\documentclass[preprint,amsmath,amssymb,superscriptaddress,showpacs]{revtex4}
\usepackage{graphicx}
\begin{document}
\renewcommand{\thesection}{\arabic{section}}
\renewcommand{\thesubsection}{\arabic{section}.\arabic{subsection}}
\renewcommand{\thefigure}{\arabic{figure}}
\baselineskip=0.7cm

\title{Valley- and spin-filter in monolayer MoS$_2$}

\author{Leyla Majidi}
\affiliation {School of Physics, Institute for Research in Fundamental Sciences (IPM), Tehran 19395-5531, Iran}
\author{Moslem Zare}
\affiliation {School of Physics, Institute for Research in Fundamental Sciences (IPM), Tehran 19395-5531, Iran}
\author{Reza Asgari~\footnote{Corresponding author: Tel: +98 21 22280692; fax: +98 21 22280415.\\ E-mail address: asgari@ipm.ir}}
\affiliation {School of Physics, Institute for Research in Fundamental Sciences (IPM), Tehran 19395-5531, Iran}

\begin{abstract}
We propose a valley- and spin-filter based on a normal/ferromagnetic/normal molybdenum disulfide (MoS$_2$) junction where the polarizations of the valley and the spin can be inverted by reversing the direction of the exchange field in the ferromagnetic region. By using a modified Dirac Hamiltonian and the scattering formalism, we find that the polarizations can be tuned by applying a gate voltage and changing the exchange field in the structure. We further demonstrate that the presence of a topological term ($\beta$) in the Hamiltonian results in an enhancement or a reduction of the charge conductance depending on the value of the exchange field.
\end{abstract}

\pacs{ }
\maketitle



\section{Introduction}
\label{Intro}
The investigation of the internal quantum degrees of freedom of electrons lies at the heart of the condensed matter physics. Interest in the electron spin, the most studied examples, extends to other quantum degrees of freedom of electrons such as a valley. The recent emergence of two-dimensional layered materials - in particular the transition metal dichalcogenides~\cite{Mattheis73} such as molybdenum disulfide (MoS$_2$) - has provided new opportunities to explore the quantum control of the valley degree of freedom.

The monolayer MoS$_2$ has recently
attracted great interest because of its potential applications in two-dimensional (2D)
nanodevices~\cite{Mak10, Radisavljevic}, owing to the structural stability and lack of dangling bonds, although it had been obtained and studied in the several
decades ago~\cite{Banerjee}. The monolayer MoS$_2$ is a direct gap semiconductor with a band gap of $1.9$ eV ~\cite{Mak10,Splendiani10,Korn11} which enables a wide range of applications such as transistors~\cite{Radisavljevic11,Fontana13} and optoelectronic devices~\cite{Wang12,Sallen12}. Similar to graphene, the conduction and valence band edges consist of two degenerate valleys $ (K, K') $ located at the corners of the hexagonal Brillouin zone. Due to the large valley separating in the momentum space, in the case of the absence of short-range interactions, the intervalley scattering~\cite{lu} should be negligible and thus the valley index becomes a new quantum number. Therefore, manipulating the valley quantum number can produce new physical effects. One of the peculiarities of a monolayer MoS$_2$ is the coupled spin-valley in the electronic structure, which is owing to the strong spin-orbit coupling (originating from the existence of a heavy transition metal in the lattice structure) and the broken inversion symmetry~\cite{Xiao12,Zeng12,Cao12}. In addition, the presence of the strong spin-orbit coupling produces a spin splitting of the valence band and makes the monolayer MoS$_2$ a convenient platform to explore new spin-dependent phenomena and their implementation in spintronic devices. These features provide us a new way to generate spin- and valley-polarized current in monolayer MoS$_2$. Recent studies stated that a ferromagnetic behavior~\cite{Zhang07,Li09,Mathew12,Ma12,Tongay,Mishra13,Vojvodic09,Ataca11} and a superconducting~\cite{Gupta91, Takagi12, Ye12, Roldan13} transition can be occurred in MoS$_2$ sheet. Also, the single-layer and multi-layer MoS$_2$ can be $n$- or $p$-type doped on generating desirable charge carriers~\cite{Radisavljevic11, Fontana13}. Recently, the properties of the charge, spin and valley transport in a monolayer molybdenum disulfide $p$/$n$/$p$~\cite{Sun14}, ferromagnetic/superconducting/ferromagnetic (F/S/F)~\cite{Majidi14_2} and normal/superconducting (N/S)~\cite{Majidi14_1} junctions have been investigated.

In this paper, we focus on the transport properties of a normal/ferromagnetic/normal (N/F/N) MoS$_2$ junction. We find, by using a modified Dirac Hamiltonian and the scattering formalism, that this structure produces the polarizations of the valley and the spin in the right N region. The polarizations can be tuned by changing the chemical potential of the N and F regions, $\mu$, by means of a gate voltage, and the exchange field of the F region, $h$. In particular, we find that the current through this junction is fully valley- and spin-polarized for a selected range of the chemical potential and furthermore, the polarizations can be inverted by reversing the direction of the exchange field. In addition, we demonstrate that the topological term ($\beta$) in the Hamiltonian of MoS$_2$ results in an enhancement or a reduction of the charge conductance, depending on the value of the exchange field.

\section{Model and Theory}
\label{Theory}
\begin{figure}[t]
\begin{center}
\includegraphics[width=3.5in]{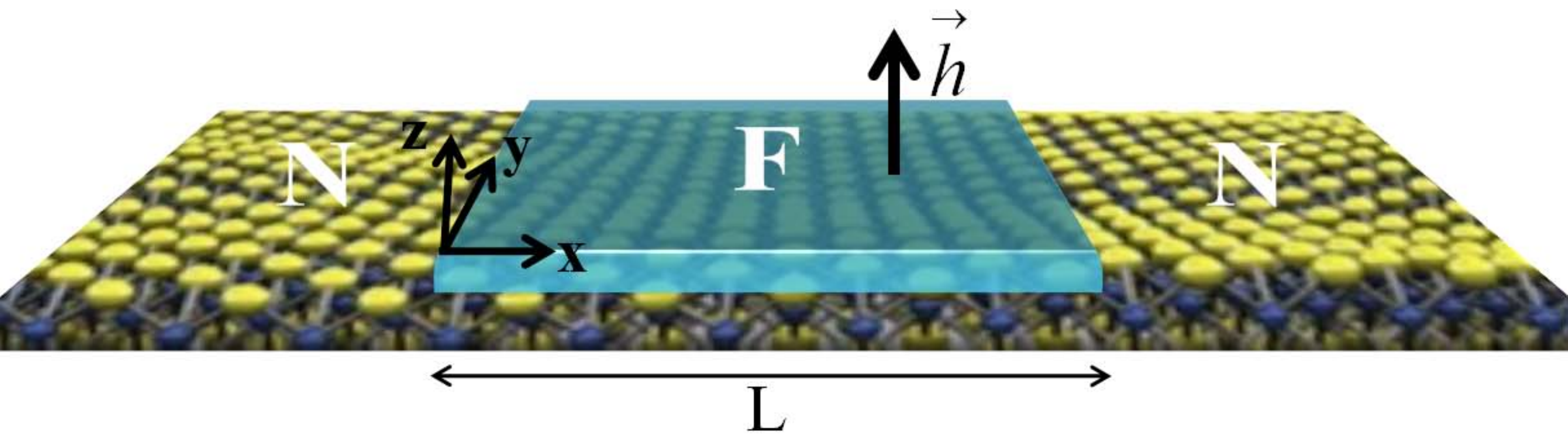}
\end{center}
\caption{\label{Fig:1} (Color online) Schematic illustration of the proposed normal/ferromagnetic/normal (N/F/N) MoS$_2$ structure with $p$-doped, N and F regions: the intermediate region is the proximity induced ferromagnet with the exchange field $\vec{h}=h\ \hat{z}$.}
\end{figure}
To study the valley- and spin-polarized quantum transport in a monolayer molybdenum disulfide (MoS$_2$), we consider a wide $p$-doped normal/ferromagnetic/normal (N/F/N) MoS$_2$ junction in the $x-y$ plane in which the F region with the exchange field $\vec{h}=h\ \hat{z}$ ($0<x<L$) connects to the two N regions ($x<0$, $x>L$), as shown schematically in Fig. \ref{Fig:1}. The ferromagnetism is assumed to be induced by means of the proximity effect to the F lead with desired properties. Such a F region in graphene can be produced by using an insulating ferromagnetic substrate, or by adding F metals or magnetic impurities on the top of the graphene sheet~\cite{Swartz12,Dugaev06,Uchoa08,Tombros07,Yazyev10}.

The low-lying electronic states in a monolayer MoS$_2$ can be described by the modified-Dirac Hamiltonian~\cite{Rostami13}
\begin{equation}
\label{H}
H=v_{\rm F}(\vec{\sigma}_{\tau}. \vec{p})+\frac{\Delta}{2}\sigma_{z}+\lambda s \tau\ (\frac{1-\sigma_z}{2})+\frac{{\vec{p}}^2}{4m_0}(\alpha+\beta\sigma_z)-h s,
\end{equation}
for spin $s=\pm 1$ and valley $\tau=\pm 1$, and contains an additional term $-hs$ in the presence of an exchange interaction in the F region. The $\alpha$ term is originated from the difference between electron and hole masses recently reported by using \textit{ab initio} calculations~\cite{Peelaers12} and in addition, the $\beta$ term leads to a new topological characteristic~\cite{rostami_opt}. The numerical values of the other parameters will be given in Section \ref{results}. The modified Hamiltonian $H$ acts on the two component spinor of the form $(\psi_c,\psi_v)$, where $c$ and $v$ denote the conduction and valence bands, respectively, and $\vec{\sigma}_{\tau}=(\tau\sigma_x,\sigma_y,\sigma_z)$
is the vector of the Pauli matrices acting on the conduction band as well as the valence band.

In order to compute the charge, spin, and valley conductances, we consider an incident electron in the left N region with spin-$s$ from valley-$\tau$. Taking into account the normal reflection with the coefficient $r_{s,\tau}$, as well as the transverse wave vector $k_y$, the total wave functions in the left N, F, and right N regions, signed by 1, 2, and 3, respectively can be written as:
\begin{eqnarray} \label{eq:psi}
\psi_{1}&=&
\frac{1}{\sqrt{A_1}}\ e^{-i\tau k_{1x} x} e^{ik_yy}
\left(
\begin{array}{c}
e^{{i\tau \theta_{1}/2 }} \\
- a_1\ \tau\ e^{{-i\tau \theta_{1} }/2}
\end{array}
\right)+\frac{r_{s,\tau}}{\sqrt{A_1}}\ e^{i\tau k_{1x} x} e^{ik_yy}
\left(
\begin{array}{c}
e^{{-i\tau \theta_{1}}/2} \\
 a_{1}\ \tau\ e^{{i\tau \theta_{1} }/2}
\end{array}
\right),\\
\psi_{2}&=&
\frac{a'_{s,\tau}}{\sqrt{A_2}}\ e^{-i\tau k_{2x} x} e^{ik_yy}
\left(
\begin{array}{c}
e^{{i\tau \theta_{2} }/2} \\
- a_{2}\ \tau\ e^{{-i\tau \theta_{2} }/2}
\end{array}
\right)+\frac{b'_{s,\tau}}{\sqrt{A_2}}\ e^{i\tau k_{2x} x} e^{ik_yy}
\left(
\begin{array}{c}
e^{{-i\tau \theta_{2} }/2} \\
 a_{2}\ \tau\ e^{{i\tau \theta_{2}/2 }}
\end{array}
\right), \\
\psi_{3}&=&
\frac{t_{s,\tau}}{\sqrt{A_{3}}}\ e^{-i\tau k_{3x} x} e^{ik_yy}
\left (
\begin{array}{c}
e^{{i\tau \theta_{3}/2}} \\
- a_ {3} \ \tau\ e^{{-i\tau \theta_{3}} /2}
\end{array}
\right).
\end{eqnarray}
Here, $a'_{s,\tau}$ and $b'_{s,\tau}$ are the coefficients of the incoming and outgoing electrons in the F region, and $t_{s,\tau}$ is the transmission amplitude of the electron in the right N region. $\theta_j=\arctan({k_y/k_{jx}})$ indicates the angle of the propagation of the electron which has the longitudinal wave vector $k_{jx}=\sqrt{{|\vec{k}_j|}^2-{k_y}^2}$ with the momentum-energy relation
\begin{eqnarray}
|\vec{k}_j|&=&[\gamma \pm \frac{2m_0}{{\hbar^2(\alpha^2-\beta^2)}} ([2 (h_j s+\mu_j+\varepsilon)-\Delta][2(h_j s
+\mu_j+\varepsilon -\lambda s \tau)+\Delta]\ (\beta^2-\alpha^2)\nonumber\\
&+&[2\alpha(h_j s+\mu_j+\varepsilon)+\beta \Delta-(\alpha+\beta)\lambda s \tau+4m_0  {v_{\rm F}} ^2] ^2) ^ {1/2}] ^ {1/2}.
\end{eqnarray}
Here, $\varepsilon$ is the excitation energy, $h_1=h_3=0$, $h_2=h$, and $\gamma=(8m^2_0 v_{\rm F}^2 +2 m_0[2\alpha(h_j s+\mu_j+\varepsilon)+\beta \Delta-\lambda s \tau(\alpha+\beta)])/{\hbar^2 (\alpha^2-\beta^2)}$. Furthermore, we define $a_j={\hbar v_{\rm F}|\vec{k}_j|}/{[\mu_j + \varepsilon+h_j s+{\Delta}/{2} -\lambda s \tau- {\hbar^2 |\vec{k}_j|^2 (\alpha- \beta)}/{4m_0}]}$ and $A_j= {\hbar |\vec{k}_j| \cos(\tau\theta_j)[\alpha+\beta+a_j^2(\alpha-\beta)]}/{4m_0 v_{\rm {F}}}+a_j\cos(\tau\theta_j)$.
The transmission, $t_{s,\tau}$ and reflection, $r_{s,\tau}$ coefficients, which allow the computation of the transport properties, can be obtained by applying the continuity of the wave functions at the two interfaces ($x=0$ and $x=L$). By defining the normalized valley and spin resolved conductances (in units of $e^2/h$):
\begin{equation} \label{eq:Gtotal}
G_{s,\tau} = \frac{1}{2} \int_{-{\pi}/{2}} ^{{\pi}/{2}}{|t_{s,\tau}|}^2 \cos{\theta_{1}}\ d\theta_{1},
\end{equation}
We calculate the charge conductance $G_c$ and the valley- and spin-polarizations, $P_v$ and $P_s$ respectively, as
\begin{eqnarray} \label{eq:gc}
G_c&=& \sum_{s=\pm 1, \tau=\pm 1}G_{s,\tau},\\
\label{eq:Pv}
P_v&=& \frac{G_K-G_{K'}}{G_K+G_{K'}},\\
\label{eq:Ps}
P_s &=& \frac{G_{\uparrow}-G_{\downarrow}}{G_{\uparrow}+G_{\downarrow}},
\end{eqnarray}
where $G_{K^{(')}}=\sum_{s=\pm 1, \tau=1(-1)}G_{s,\tau}$ and $G_{\uparrow(\downarrow)}=\sum_{\tau=\pm 1, s=1(-1)}G_{s,\tau}$ are the valley- and spin-resolved conductances, respectively.
\section{Numerical results and discussions}
\label{results}
\begin{figure}[t]
\begin{center}
\includegraphics[width=3.5in]{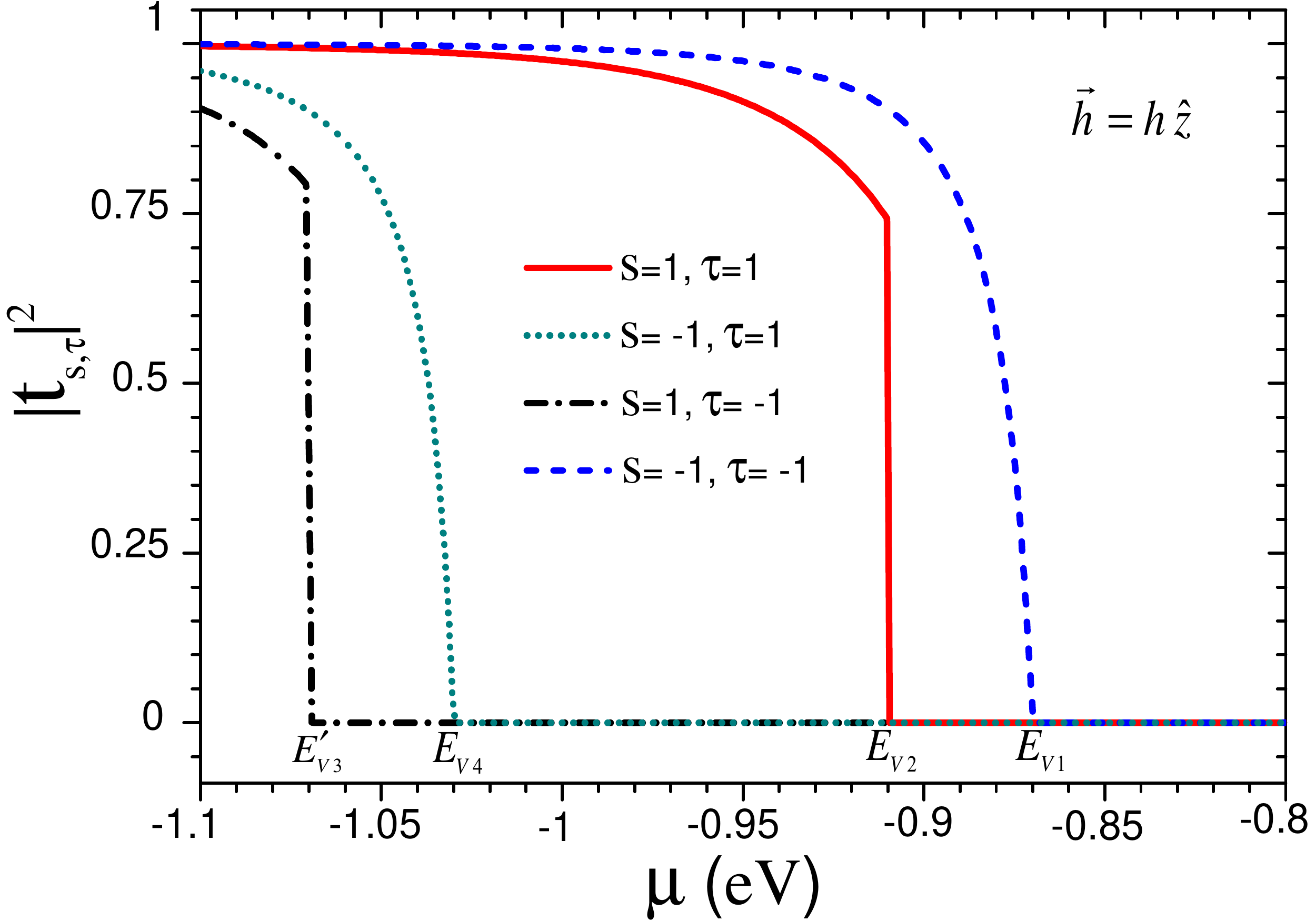}
\end{center}
\caption{\label{Fig:2} (Color online) The transmission probability of an incident electron with spin-$s$ ($s=\pm 1$) from valley-$\tau$ ($\tau=\pm 1$) of the left N region as a function of the chemical potential $\mu$, for normal incidence ($\theta_1=0$) to the MoS$_2$-based N/F/N structure with the exchange field $\vec{h}=h\ \hat{z}$ for the F region, when $\alpha=0.43$, $\beta=2.21$, $h/\lambda=0.5$, and $L/a=10$.}
\end{figure}

In order to evaluate the numerical results using the numerical transmission amplitude $t_{s,\tau}$, and Eqs. (\ref{eq:gc}), (\ref{eq:Pv}), and (\ref{eq:Ps}) at zero temperature, we set the energy gap $\Delta= 1.9$ eV, the spin-orbit coupling constant $\lambda= 0.08$ eV, the Fermi velocity $v_{\rm F}=0.53\times10^6$ m/s, $\alpha = 0.43$, and $\beta=2.21$. We take the chemical potentials  $\mu_1=\mu_2=\mu_3=\mu$ ($\mu<0$). The energies $\mu$, $\Delta$, and $\lambda$ are in units of electron volts, eV. The exchange field and the length of the F region are in units of the spin-orbit coupling ($\lambda$) and lattice ($a$) constants, respectively. Also, we set $\varepsilon=eV=0$ at zero temperature.

\begin{figure}[t]
\begin{center}
\includegraphics[width=3.2in]{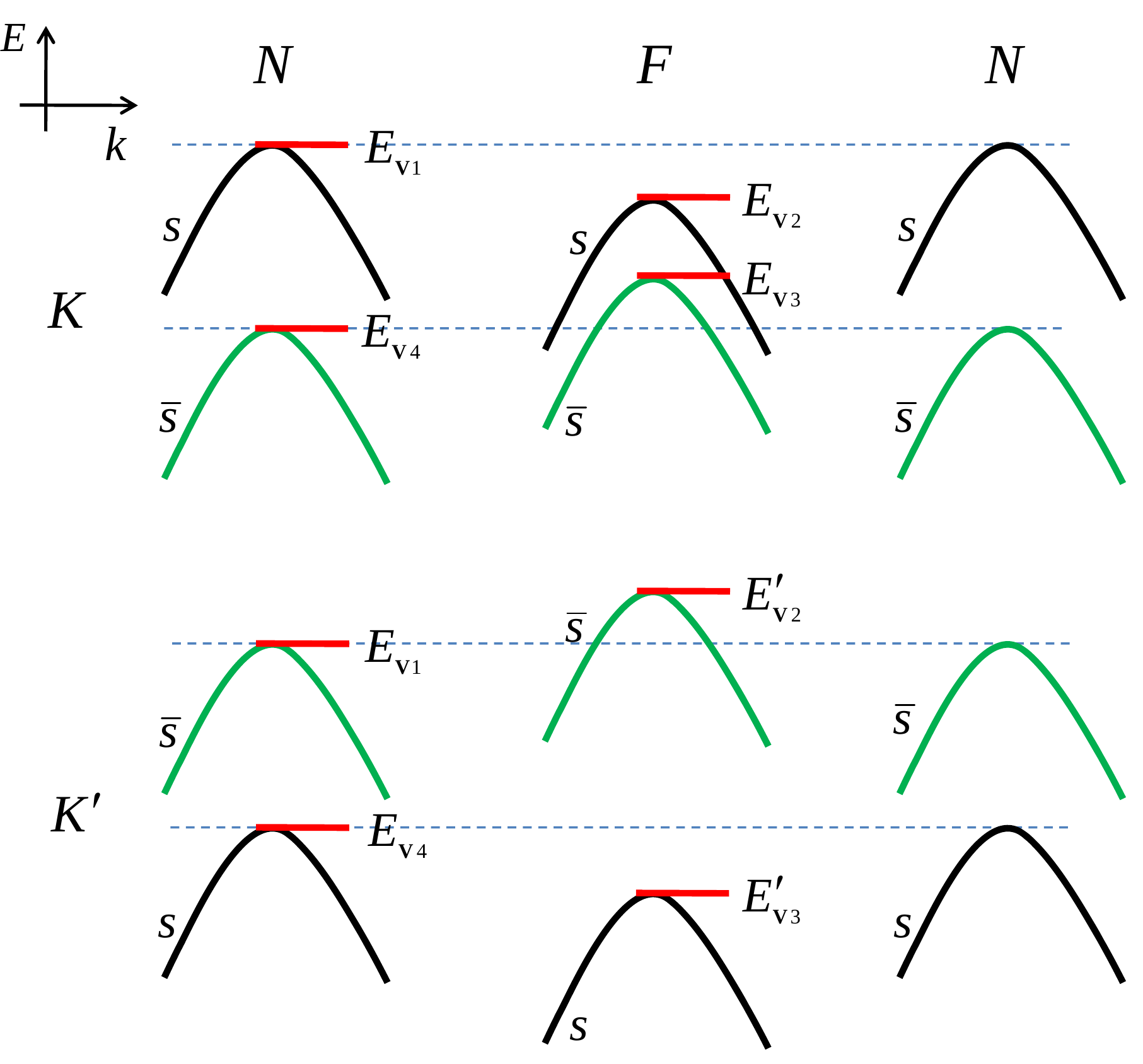}
\end{center}
\caption{\label{Fig:3} (Color online) The dispersion relation in momentum space of the N and F regions of the proposed N/F/N structure with the exchange field $\vec{h}=h\ \hat{z}$ ($h<\lambda$), near the two $K$ ($\tau=1$) and $K'$ ($\bar{\tau}=-\tau$) valleys. The energy of the valence band edges for different spin-subbands ($s=1$, $\bar{s}=-s$) near the two valleys of N and F regions are $E_{V1}=-\Delta/2+\lambda$, $E_{V2}=-\Delta/2+\lambda-|h|$, $E'_{V2}=-\Delta/2+\lambda+|h|$, $E_{V3}=-\Delta/2-\lambda+|h|$, $E'_{V3}=-\Delta/2-\lambda-|h|$, and $E_{V4}=-\Delta/2-\lambda$, which are measured from the center of the gap $\Delta$ (zero-energy point).}
\end{figure}
Fig. \ref{Fig:2} shows the transmission probability of a normal incident electron with spin-$s$ from valley-$\tau$ of the left N region in terms of the chemical potential $\mu$, when the length of the F region is $L/a=10$. $s=\pm 1$ denotes the up- (down-) spin and $\tau=\pm1$ denotes the two independent valleys ($K$ and $K'$) in the first Brillouin zone. It is seen that for each set of $(s,\tau)$, there is a critical chemical potential above which the transmission of the corresponding electron suppresses. Most remarkably, we observe that the transport of charge is governed purely by the incident electrons with spin-down ($s=-1$) from $K'$ valley ($\tau=-1$), when $E_{V2}<\mu\leq E_{V1}$. This behavior can be understood by the band structures near the $K$ and $K'$ valleys of the $p$-doped N and F regions of the proposed structure with the exchange field $\vec{h}=h\ \hat{z}$ (see Fig. \ref{Fig:3}). The energies $E_ {Vi} $ and $E'_ {Vi} $ ($i=1-4$) define the energy of the valence band edges for different spin-subbands of two valleys of the N and F regions. As seen from Fig. \ref{Fig:3}, the presence of the exchange field in the F region shifts the spin-up ($s=1$) subband of the $K$ ($K'$) valley down- (up-) ward, and the spin-down ($\bar{s}=-s$) subband of the $K$ ($K'$) valley up- (down-) ward by $|h|$. Thus for $E_{V2}<\mu\leq E_{V1}$ ($E_{V1}=-\Delta/2+\lambda$, $E_{V2}=-\Delta/2+\lambda-|h|$), the incident electrons with spin-up from $K$ valley ($\tau=1$) are filtered and the transport of electrons with $s=\tau=-1$ leads to the fully valley- and spin-polarized charge current while for $\mu\leq E_{V2}$, the incident electrons from different spin-subbands of two valleys contribute to the charge current in the proposed N/F/N structure. Furthermore, we find that the type of the valley- and spin-polarizations of the transmitted electron with $E_{V2}<\mu\leq E_{V1}$ can be changed by reversing the direction of the exchange field in F region. This can be understood from the band structures in Fig. \ref{Fig:3} by changing the sign of the spin ($s\leftrightarrow\bar{s}$) and valley ($\tau\leftrightarrow\bar{\tau}$) degrees of freedom.

\begin{figure}[]
\begin{center}
\includegraphics[width=6in]{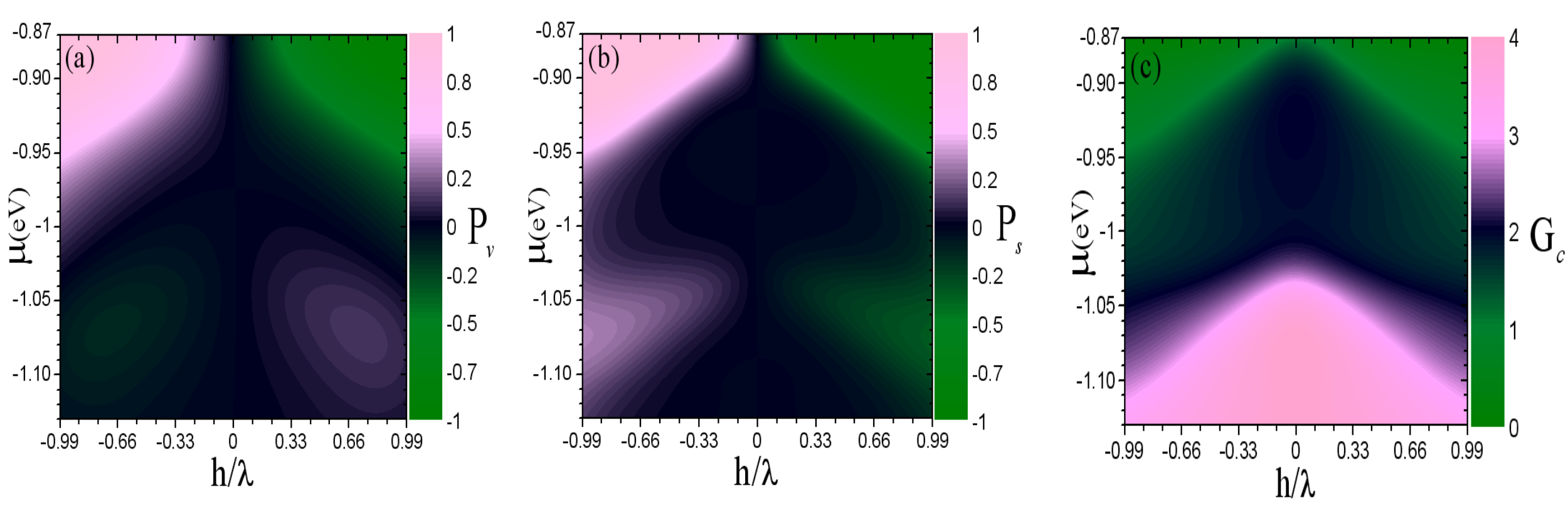}
\end{center}
\caption{\label{Fig:4} (Color online) Plots of the valley-polarization $P_v$, spin-polarization $P_s$, and the charge conductance $G_c$ as a function of $h/\lambda$ and $\mu$, when $\alpha=0.43$, $\beta=2.21$, and $L/a=10$.}
\end{figure}

The resulting charge conductance $G_c$ and its valley- and spin-polarizations, $P_v$ and $P_s$ respectively, are presented in Fig. \ref{Fig:4} in terms of the ratio $h/\lambda$ and the chemical potential $\mu$, when the length of the F region is $L/a=10$. It is seen that $P_v$, $P_s$, and $G_c$ change significantly by varying the chemical potential $\mu$ and the exchange field $h$. The charge conductance has full valley- and spin-polarizations for a selected range of the $\mu$ and $h$ ($E_{V2}<\mu\leq E_{V1}$), and also relatively small polarizations ($\leq 0.5$) for a wide parameter region ($E'_{V3}<\mu\leq E_{V4}$) with the same sign for $P_s$ and the opposite sign for $P_v$. In addition, we find that the polarizations, $P_v$ and $P_s$, are odd with respect to the exchange field such that the type of the polarizations can be changed by changing the sign of the exchange field.
\begin{figure}[t]
\begin{center}
\includegraphics[width=3.5in]{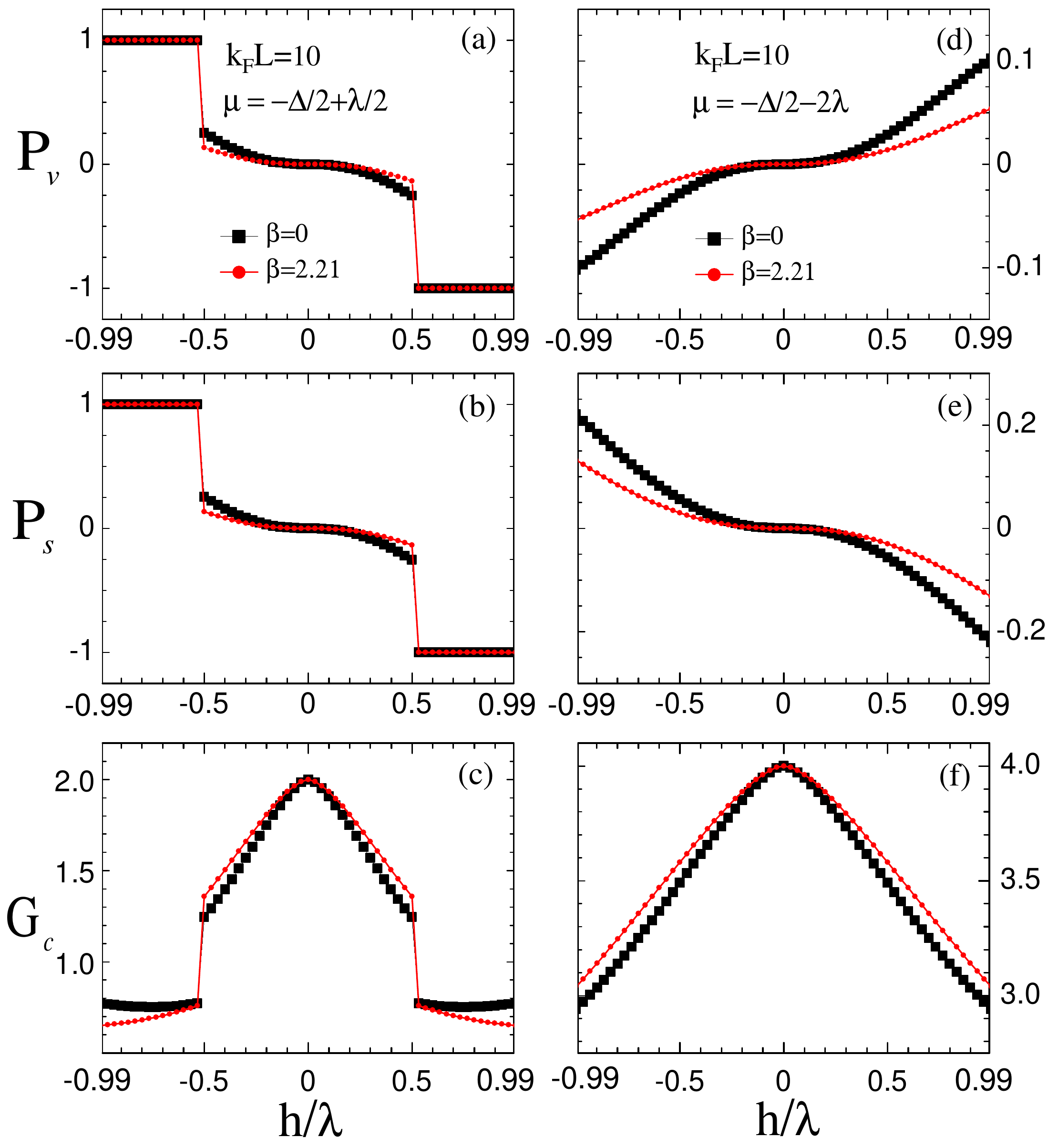}
\end{center}
\caption{\label{Fig:5} (Color online) Valley-polarization $P_v$, spin-polarization $P_s$, and the charge conductance $G_c$ versus $h/\lambda$ for two values of the chemical potential $\mu=-\Delta/2+\lambda/2$ (left panel) and $\mu=-\Delta/2-2\lambda$ (right panel), when $k_FL=10$, $\alpha=0.43$ and $\beta=0$ and $2.21$.}
\end{figure}

Furthermore, the behavior of the polarizations $P_v$ and $P_s$, and the charge conductance $G_c$ versus $h/\lambda$ are shown in Fig. \ref{Fig:5} for two values of the chemical potential in the absence and presence of the topological term ($\beta$), when $\alpha=0.43$ and $k_{F}L=10$ ($k_{F}=\mu/\hbar v_{\rm F}$). Depending on the value of $\mu$, the charge conductance $G_c$ has full valley- and spin-polarizations for the exchange fields in the range $|h|>|\mu|-\Delta/2+\lambda$ (left panel) and small polarizations for $|h|\leq|\mu|-\Delta/2+\lambda$ (left and right panels). Also, the magnitude of $G_c$ decreases by increasing the magnitude of the exchange field, $|h|$. More importantly, we see that the presence of the $\beta$ term in the modified Hamiltonian can enhance or reduce $G_c$, depending on the $h/\lambda$ value, and reduces the value of the polarizations for $|h|\leq|\mu|-\Delta/2+\lambda$. Moreover, we find that the presence of the mass asymmetry term ($\alpha$) in the Hamiltonian of MoS$_2$ has no significant effect on the valley- and spin-resolved conductances (similar behaviors have been demonstrated for local and nonlocal AR processes in
Refs. \cite{Majidi14_2,Majidi14_1}).
\section{Conclusion}
\label{Conclusion}
In conclusion, we have investigated the realization of the valley- and spin-polarized quantum transport in a monolayer molybdenum disulfide (MoS$_2$). We have demonstrated that a normal/ferromagnetic/normal (N/F/N) MoS$_2$ junction, is capable of producing a highly population of the valleys and the spin. In particular, the charge current through this junction has full valley- and spin-polarizations if the chemical potential of the N and F regions lies in a determined region of the parameters and a small polarization ($\leq 0.5$) for a wide parameter region. Moreover, the polarity of the proposed valley- and spin-filter can be inverted by reversing the direction of the exchange field in the F region. Furthermore, we have found that the additional topological term ($\beta$) in the Hamiltonian of MoS$_2$ results in an enhancement or a reduction of the charge conductance, depending on the value of the exchange field. We note that the role of the finite-size effect for a nanoribbon MoS$_2$ has not been addressed in the present work.








\end{document}